\documentclass[a4paper,fleqn,usenatbib]{mnras}
\usepackage[T1]{fontenc}
\usepackage{ae,aecompl}

\usepackage{graphicx}   
\usepackage{amsmath}    
\usepackage{amssymb}    

\newcommand{\figureref}[1]{Fig.~\ref{#1}}

\newcommand{\sectionref}[1]{Section \ref{#1}}

\title[Stability of supersonic ionized accretion flows]{Testing the
stability of supersonic ionized Bondi accretion flows with radiation
hydrodynamics}

\author[B. Vandenbroucke et al.]{
Bert Vandenbroucke,$^{1}$\thanks{E-mail~: bv7@st-andrews.ac.uk}
Nina S. Sartorio,$^{2}$
Kenneth Wood,$^{1}$
Kristin Lund,$^{1}$
\newauthor
Diego Falceta-Gon\c{c}alves,$^{2,3}$
Thomas J. Haworth,$^{4}$
Ian Bonnell,$^{1}$
Eric Keto,$^{5}$
\newauthor
Daniel Tootill,$^{1}$
\\
\\
$^1$SUPA, School of Physics \& Astronomy, University of St Andrews, North Haugh,
St Andrews, KY16 9SS, United Kingdom
\\
$^2$Instituto Nacional de Pesquisas Espaciais - INPE, Divis\~{a}o de
Astrof\'{i}sica, Av. dos Astronautas, 1.758 - S\~{a}o Jos\'{e} dos Campos - SP,
Brazil
\\
$^3$Escola de Artes, Ci\^{e}ncias e Humanidades, Universidade de S\~{a}o Paulo,
Rua Arlindo Bettio 1000, S\~{a}o Paulo, SP, 03828-000, Brazil
\\
$^4$Astrophysics Group, Imperial College London, Blackett Laboratory,
Prince Consort Road, London SW7 2AZ, UK
\\
$^5$Harvard-Smithsonian Center for Astrophysics, Cambridge, MA 02138, USA
}

\date{Accepted XXX. Received YYY; in original form ZZZ}

\pubyear{2019}

\begin{document}

\label{firstpage}
\pagerange{\pageref{firstpage}--\pageref{lastpage}}
\maketitle

\begin{abstract}
We investigate the general stability of 1D spherically symmetric ionized Bondi
accretion onto a massive object in the specific context of accretion onto a
young stellar object. We first derive a new analytic expression for a
steady state two temperature solution that predicts the existence of compact
and hypercompact H\textsc{ii} regions. We then show that this solution is only
marginally stable if ionization is treated self-consistently. This leads to a
recurring collapse of the H\textsc{ii} region over time. We derive a
semi-analytic model to explain this instability, and test it using spatially
converged 1D radiation hydrodynamical simulations. We discuss the implications
of the 1D instability on 3D radiation hydrodynamics simulations of supersonic
accreting flows.
\end{abstract}


\section{Introduction}

The problem of a spherically symmetric accretion flow onto a massive 
star including a rigorous treatment of the hydrodynamics was first 
introduced by \citet{1952Bondi}, and is consequently referred to as the 
Bondi problem. \citet{1954Mestel} studied the effect of ionizing 
radiation on the original Bondi problem, using an approximation whereby 
both the ionized and the neutral region are assumed to be isothermal, 
with different isothermal sound speeds and a sharp transition from 
ionized to neutral regime at the ionization front. \citet{2002Keto, 
2003Keto} showed that this work explains the existence of a steady state 
solution in which an ionized region is trapped inside the accreting 
flow. This explains the existence of trapped compact and ultracompact 
H\textsc{ii} regions around massive accreting stars. These ultracompact 
H\textsc{ii} regions are indeed observed around massive protostars, with 
observed size estimates varying from $\sim{}10^3$~AU 
\citep{2002Churchwell} to as small as $\sim{}10$~AU \citep{2016Ilee}. He 
also showed how accretion can continue while the protostar is already 
emitting UV radiation, which has implications for the final mass of a 
young stellar object.

Recently, this work was extended to 3D using a Monte Carlo radiation 
hydrodynamics code (Lund \emph{et al.}, submitted to MNRAS). They 
were able to numerically simulate a trapped H\textsc{ii} region, and 
also showed the change in regime from a trapped R type ionization front 
into an expanding D type ionization front under a changing source 
luminosity, with the subsequent shut down of accretion onto the central 
star. However, their work also casted doubt about the stability of the 
steady state solution of \citet{2002Keto}, as their trapped H\textsc{ii} 
regions showed signs of periodic collapse, whereby the ionization front 
periodically collapsed onto the central star and moved outwards again.

In this work, we will analyse this instability in more detail, using a 
high resolution 1D spherically symmetric radiation hydrodynamics code, 
and a semi-analytic stability analysis.

Note that massive stars are expected to form through an accretion disk 
\citep{2016Beltran, 2018Kuiper}, which leads to more extended 
ultracompact H\textsc{ii} regions, and stabilises the ionization front 
(Sartorio \emph{et al}, submitted to MNRAS). This renders the assumption 
of spherical symmetry an unlikely scenario for accretion onto a 
protostar. Nevertheless, this work is still relevant for our 
understanding of trapped H\textsc{ii} regions and can serve as a useful 
benchmark test for models that combine hydrodynamics, photoionization 
and gravitational accretion.

\section{Analytics}
\label{section:analytics}

In this section, we present the analytical solution to the original 
Bondi problem, and the extended Bondi problem with ionizing radiation. 
This work is a summary of the original work of \citet{1952Bondi} on 
steady state accretion, and extends the work of \citet{1954Mestel} and 
\citet{2002Keto} with a full analytic expression for two temperature 
steady state accretion. We will summarize the main equations and present 
the full analytic steady state solution, before addressing the stability 
of this analytic solution under self-consistent ionization.

\subsection{Bondi accretion}

We assume spherically symmetric flow onto a massive object of constant mass $M$
with a constant accretion rate \citep{1952Bondi}
\begin{equation}
\dot{M}_a = 4\pi{}r^2\rho{}(r) v(r),
\label{eq:accretion_rate}
\end{equation}
with $\rho{}(r)$ and $v(r)$ the mass density and fluid velocity as a function of
radius $r$.

If we furthermore neglect the gravitational force due to the mass of the 
accreting fluid and assume the potential is dominated by $M$, and assume 
an isothermal equation of state with a constant sound speed $c_s(r) = 
\sqrt{{\rm{}d}P(r) / {\rm{}d}\rho{}(r)} = C_s$, the constancy of the 
accretion rate leads to a specific form of Bernoulli's equation:
\begin{equation}
\frac{v^2(r)}{2C_s^2} + \ln{}(\rho{}) - \frac{2R_B}{r} = {\rm{}constant},
\label{eq:Bernoulli_general}
\end{equation}
where we introduced the (constant) Bondi radius
\begin{align}
r_B(r) = \frac{GM}{2c_s^2(r)} &= R_B \\
&\approx{}110.9 
\left(\frac{M}{{\rm{}M}_\odot{}}\right) 
\left(\frac{C_s}{2~{\rm{}km~s}^{-1}}\right)^{-2}~{\rm{}AU},
\end{align}
with $G=6.674\times{}10^{-11}~{\rm{}m}^3~{\rm{}kg}^{-1}~{\rm{}s}^{-2}$ 
Newton's constant.

Assuming that the velocity vanishes at infinity, we find the value of 
the constant:
\begin{equation}
\frac{v^2(r)}{2C_s^2} + \ln{}(\rho{}) - \frac{2R_B}{r} = \ln{}(\rho{}_\infty{}),
\label{eq:Bernoulli}
\end{equation}
with $\rho{}_\infty{}$ the rest density of the accreting material. By 
evaluating this equation at $r = R_B$, we can find the density 
$\rho{}_B$ at the Bondi radius: $\rho{}_B = e^{3/2} \rho{}_\infty{} 
\approx{} 4.48 \rho{}_\infty{}$.

Using (\ref{eq:accretion_rate}), we find an expression for the density as a
function of radius:
\begin{equation}
\rho{}(r) = -\frac{\rho{}_B R_B^2 C_s}{r^2 v(r)}.
\label{eq:Bondi_density}
\end{equation}
Substituting this in (\ref{eq:Bernoulli}), we find
\begin{equation}
\left(\frac{v(r)}{C_s}\right)^2 -
\ln{}\left[\left(\frac{v(r)}{C_s}\right)^2\right] -
4\ln{}\left(\frac{r}{R_B}\right) + 3 - 4\frac{R_B}{r} = 0,
\end{equation}
which has general solution \citep{2004Cranmer}
\begin{equation}
v(r) = -C_s \sqrt{-W\left(-\left(\frac{R_B}{r}\right)^4 \exp\left(3 -
4\frac{R_B}{r}\right) \right)},
\end{equation}
where $W(x)$ is the Lambert \emph{W} function, and the negative root was chosen
since we want material to be accreting onto the central object.

The Lambert \emph{W} function is a complex valued function with an 
infinite number of branches, but two of these branches, $W_0(x)$ and 
$W_{-1}(x)$, are real valued in the range $[-{1}/{e}, 0]$, with 
$W_0\left(-{1}/{e}\right) = W_{-1}\left(-{1}/{e}\right) = -1$, $W_0(0) = 
0$, and $W_{-1}(x) \rightarrow{} -\infty{}$ for $x \rightarrow{} 0$. 
Since we want the velocity to be a strictly increasing function of 
radius, we see that we start on the $W_0$ branch for $r > R_B$, and 
switch to the $W_{-1}$ branch at $r = R_B$. For $r \rightarrow{} 0$, the 
velocity diverges.

We end up with the following analytic expressions for the velocity for steady
state accretion onto a massive object:
\begin{equation}
v(r) = \begin{cases}
-C_s \sqrt{-W_{-1}\left(-\omega{}(r)\right)} & r \leq{} R_B,\\
-C_s \sqrt{-W_0\left(-\omega{}(r)\right)} & r > R_B,
\label{eq:Bondi_velocity}
\end{cases}
\end{equation}
with
\begin{equation}
\omega{}(r) = \left(\frac{R_B}{r}\right)^4 \exp\left(3 -
4\frac{R_B}{r}\right).
\end{equation}

\subsection{Ionizing radiation}
\label{subsection:ionizing_radiation}

To include ionization from a central isotropic point source, we follow
\citet{1954Mestel} and assume a sharp transtion
from ionized to neutral at a well defined radius $R_I$. Both inside and outside
the ionized region the gas obeys an isothermal equation of state, with the
constant sound speed given by $C_{s,i}$ and $C_{s,n}$ respectively. We will
characterize the change in equation of state in terms of the \emph{pressure
contrast} $P_c = {C_{s,i}^2}/{C_{s,n}^2}$. Since the temperature in the
ionised region will be higher than that in the neutral region, we always have
$C_{s,i} > C_{s,n}$, and hence $P_c > 1$.

Since (\ref{eq:Bondi_density}) and (\ref{eq:Bondi_velocity}) are completely
determined by the boundary condition at $r \rightarrow{} +\infty{}$, this
solution is still valid for $r \geq{} R_I$, with $C_s = C_{s,n}$, and $R_B =
R_{B,n} = {GM}/(2C_{s,n}^2)$. To find the solution for $r < R_I$, we impose
conservation of mass and momentum across the ionization front to find the
following Rankine-Hugoniot conditions:
\begin{align}
\rho{}_i(R_I) v_i(R_I) &= \rho{}_n(R_I) v_n(R_I) \\
\rho{}_i(R_I) \left( v_i^2(R_I) + C_{s,i}^2 \right) &= \rho{}_n(R_I) \left(
v_n^2(R_I) + C_{s,n}^2 \right),
\end{align}
where $\rho{}_i(r)$ and $v_i(r)$, and $\rho{}_n(r)$ and $v_n(r)$ are the density
and velocity in the ionized and neutral regions respectively.

If we introduce the \emph{jump factor}
\begin{equation}
\Gamma{} =
\frac{\rho{}_i(R_I)}{\rho{}_n(R_I)} = \frac{v_n(R_I)}{v_i(R_I)},
\end{equation}
we find the
following equation for the jump as a function of the known neutral velocity
$v_n(R_I)$:
\begin{equation}
C_{s,i}^2 \Gamma{}^2 - \left(v_n^2(R_I) + C_{s,n}^2\right) \Gamma{} + v_n^2(R_I)
= 0,
\end{equation}
with solutions
\begin{multline}
\Gamma{} = \frac{1}{2C_{s,i}^2} \left( v_n^2(R_I) + C_{s,n}^2 \right.\\
\pm{} \left. \sqrt{ \left( v_n^2(R_I) + C_{s,n}^2 \right)^2 - 4 C_{s,i}^2
v_n^2(R_I)} \right).
\label{eq:Gamma}
\end{multline}

$\Gamma{}$ only has real solutions for $|v_n(R_I)| \geq{} v_R$ or $|v_n(R_I)|
\leq{} v_D$ (note that $v_n(r)$ is always negative because we are in an
accretion regime), with
\begin{align}
v_R &= C_{s,i} \left(1 + \sqrt{1 - \left(\frac{C_{s,n}}{C_{s,i}}\right)^2}
\right), \\
v_D &= C_{s,i} \left(1 - \sqrt{1 - \left(\frac{C_{s,n}}{C_{s,i}}\right)^2}
\right),
\end{align}
which are commonly referred to as the R type and D type solutions. We will limit
ourselves to R type solutions. In the region between the R type and D type
solutions there is no solution with a single shock at the position of the
ionization front, but more complex double front solutions could be possible; we
do not study these in this work.

It is obvious that $|v_n(R_I)| > C_{s,i}$, so that the R type solution implies
supersonic neutral gas at the ionization front. Furthermore, $\Gamma{} > 1$ for
$|v_n(R_I)| \geq{} v_R$, which implies $|v_i(R_I)| < |v_n(R_I)|$.

$\Gamma{}$ has two solutions in the R type region, corresponding to a 
large and a small jump (corresponding to the positive and negative sign 
in \eqref{eq:Gamma}). We call these the \emph{weak} and \emph{strong} R 
type solution. If we assume a density structure that initially evolved 
through Bondi accretion without ionization, then the solution will 
naturally evolve into the weak solution with the smallest jump, and this 
is the solution we will assume below. In this case, the velocity in the 
ionized region will also be supersonic (see Appendix 
\ref{appendix:Gamma} for a detailed analysis of weak and strong R type 
solutions and imposed velocity restrictions).

Inside the ionized region, equation (\ref{eq:accretion_rate}) still gives us the
density:
\begin{equation}
\rho{}_i(r) = \frac{\rho{}_i(R_I)R_I^2v_i(R_I)}{r^2v_i(r)}.
\end{equation}
The Bernoulli equation (\ref{eq:Bernoulli_general}) now becomes
\begin{multline}
\frac{v_i^2(r)}{2C_{s,i}^2} + \ln{}(\rho{}_i(r)) - \frac{GM}{C_{s,i}^2 r} =\\
\frac{v_i^2(R_I)}{2C_{s,i}^2} + \ln{}(\rho{}_i(R_I)) - \frac{GM}{R_I C_{s,i}^2},
\end{multline}
since we now need to apply boundary conditions at $r = R_I$ instead of $r
\rightarrow{} +\infty{}$.

Similarly as before, we can combine this equation with (\ref{eq:Bondi_density})
to get
\begin{multline}
\left(\frac{v_i(r)}{C_s}\right)^2 -
\ln{}\left[\left(\frac{v_i(r)}{C_s}\right)^2\right] -
4\ln{}\left(\frac{r}{R_I}\right) +
\ln{}\left(\frac{v_i^2(R_I)}{C_{s,i}^2}\right) \\- 4 \frac{R_{B,i}}{r} -
\frac{v_i^2(R_I)}{C_{s,i}^2} + 4\frac{R_{B,i}}{R_I} = 0,
\end{multline}
where we introduced the Bondi radius for the ionized region $R_{B,i} = 
{GM}/(2C_{s,i}^2)$. This equation has the general solution
\begin{equation}
v_i(r) = -C_{s,i} \sqrt{-W(-\omega{}_i(r))},
\end{equation}
with
\begin{multline}
\omega{}_i(r) = \left(\frac{R_I}{r}\right)^4
\left(\frac{v_i(R_I)}{C_{s,i}}\right)^2 \\\exp\left(4\frac{R_{B,i}}{R_I} -
4\frac{R_{B,i}}{r} - \frac{v_i^2(R_I)}{C_{s,i}^2}\right).
\end{multline}

Note that as in the neutral case, this solution will switch from the $W_0$ to
the $W_{-1}$ branch at a critical radius $R_c$, for which $v_i(R_c) = -C_{s,i}$.
However, we know that this cannot happen for a weak R front, as
$v_i(r) < -C_{s,i}$. We hence only need to consider the $W_{-1}$ branch.

\begin{figure}
\centering{}
\includegraphics[width = 0.48\textwidth]{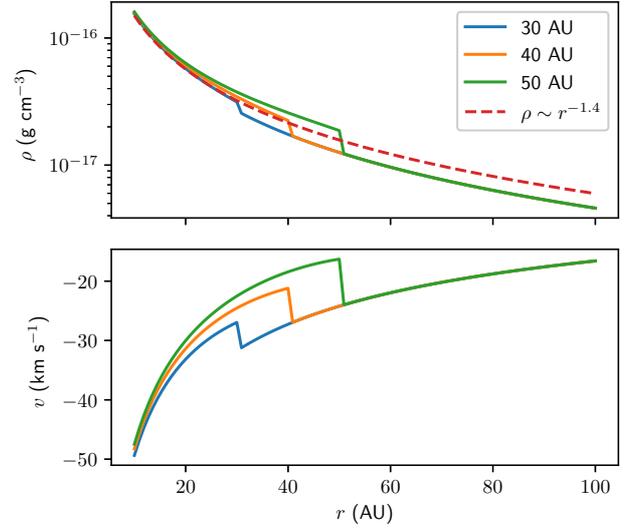}
\caption{Density (\emph{top}) and velocity (\emph{bottom}) as a function of
radius for the analytic two temperature solution for three different values of
the ionization radius $R_I$, as indicated in the legend. All models used
a central mass $M = 18~{\rm{}M}_\odot{}$, a pressure contrast $P_c=32$, a
Bondi density $\rho{}_B=10^{-19}~{\rm{}g~cm}^{-3}$ and a neutral sound speed
$C_{s,n}=2.031~{km~s}^{-1}$. The dashed line on the top panel shows a power law
approximation for the density profile in the ionized region.
\label{figure:two_temperature_solution}}
\end{figure}

The full solution for spherically symmetric accretion with ionization (in the
case $|v_n(R_I)| \geq{} v_R$) is:
\begin{align}
\rho{}(r) &= \begin{cases}
\rho{}_n(r) & r > R_I,\\
\rho{}_i(r) & r < R_I,
\end{cases}\\
v(r) &= \begin{cases}
v_n(r) & r > R_I,\\
v_i(r) & r < R_I,
\end{cases}
\end{align}
with
\begin{align}
\rho{}_n(r) &= -\frac{\rho{}_{B,n}R_{B,n}^2C_{s,n}}{r^2v_n(r)},\\
\rho{}_i(r) &= \frac{\rho{}_i(R_I)R_I^2v_i(R_I)}{r^2v_i(r)},
\end{align}
and
\begin{align}
v_n(r) &= \begin{cases}
-C_{s,n} \sqrt{-W_{-1}\left(\omega{}_n(r)\right)} & r \leq{} R_{B,n},\\
-C_{s,n} \sqrt{-W_0\left(\omega{}_n(r)\right)} & r > R_{B,n},
\end{cases}\\
v_i(r) &= -C_{s,i} \sqrt{-W_{-1}\left(\omega{}_i(r)\right)}
\end{align}
with
\begin{align}
\omega{}_n(r) = &-\left(\frac{R_{B,n}}{r}\right)^4 \exp\left(
3 - 4\frac{R_{B,n}}{r}\right),\\
\omega{}_i(r) = &-\left(\frac{R_I}{r}\right)^4
\left(\frac{v_i(R_I)}{C_{s,i}}\right)^2 \\ &\exp\left(4\frac{R_{B,i}}{R_I} -
4\frac{R_{B,i}}{r} - \frac{v_i^2(R_I)}{C_{s,i}^2}\right).
\end{align}
The parameters $\rho{}_i(R_I)$ and $v_i(R_I)$ are given by the jump conditions:
\begin{align}
\rho{}_i(R_I) &= \Gamma{} \rho{}_n(R_I),\\
v_i(R_I) &= \frac{v_n(R_I)}{\Gamma{}},
\end{align}
with
\begin{equation}
\Gamma{} = \frac{1}{2} \left( \frac{v_n^2(R_I)}{C_{s,i}^2} + \frac{1}{P_c} -
\sqrt{\left(\frac{v_n^2(R_I)}{C_{s,i}^2} + \frac{1}{P_c} \right)^2 - 4
\frac{v_n^2(R_I)}{C_{s,i}^2}} \right),
\end{equation}
so that the entire solution is parametrised in terms of $P_c$, $R_I$,
$\rho{}_{B,n}$, $C_{s,n}$ and $M$.

The density and velocity profile of the two temperature solution is 
plotted in \figureref{figure:two_temperature_solution} for three 
different values of the ionization radius $R_I$. It can be seen that 
higher values of $R_I$ lead to higher densities and lower velocities 
inside the ionised region; the solution in the neutral region is the 
same for all models.

\subsection{Self-consistent ionization}

The derivation above did not make any assumptions about how the ionization
happens, and just assumed ionization leads to a constant ionization radius
$R_I$. However, in general, the ionization radius $R_I(t)$ will vary with time,
and the time evolution will be linked to the hydrodynamical quantities,
$\rho{}(r, t)$ and $v(r, t)$. To find out if the steady-state solution obtained
above is actually \emph{stable}, we have to include this effect.

For simplicity, we will assume that the dynamical time scale $t_{\rm{}dyn}$ over
which the hydrodynamical quantities evolve is much larger than the recombination
time scale, $t_{\rm{}rec} = \left(n_{\rm{}H} \alpha{}_B(T)\right)^{-1}$, where
$n_{\rm{}H}$ is the number density of hydrogen atoms and $\alpha{}_B(T)$ is the
collisional recombination rate of ionized hydrogen to all excited levels of
neutral hydrogen, but excluding ionizations to the ground state (the so-called
on-the-spot approximation), which depends on the temperature $T$. Under this
assumption, the ionization structure of the system
instanteneously adapts to any change in density, and we can model ionization by
simply changing the equation of state.

To get the ionization radius $R_I$, we use a Str\"{o}mgren approximation, and
assume that all material inside the ionization region is completely ionized. At
the ionization radius, we have a strong discontinuity, and outside the
ionization radius the material is completely neutral. In this case, the
ionization balance equation is given by \citep{2006Osterbrock}
\begin{equation}
Q(t) = 4 \pi{} \int_{R_c}^{R_I(t)} n_H^2(r, t) \alpha{}_B(T) r^2 {\rm{}d}r,
\end{equation}
where $Q(t)$ is the total ionizing luminosity of the star. Since the steady
state density profile is very centrally peaked, and since we do not
realistically expect the profile to keep up until $r=0$, we also introduced an
inner cut off radius $R_c$, below which we assume no more ionizations to occur.

Under the two temperature approximation, the recombination rate
$\alpha{}_B(T_i) = \alpha{}_i$ is constant.
If we further assume our gas to be composed of hydrogen only, we can rewrite the
ionization balance equation as
\begin{equation}
Q(t) = \frac{4 \pi{} \alpha{}_i}{m_{\rm{}H}^2} \int_{R_c}^{R_I(t)} \rho{}^2 (r,
t) r^2 {\rm{}d}r,\label{equation:ionization_balance}
\end{equation}
where $m_{\rm{}H} = 1.674\times{}10^{-27}~{\rm{}kg}$ is the hydrogen atomic
mass. Taking the time derivative of this equation, we find the following
equation for the time evolution of $R_I(t)$:
\begin{multline}
\frac{{\rm{}d}R_I(t)}{{\rm{}d}t} = \frac{1}{\rho{}^2\left(R_I(t), t\right)
R_I^2(t)} \\ \left[ \frac{m_{\rm{}H}^2}{4 \pi{} \alpha{}_i}
\frac{{\rm{}d}Q(t)}{{\rm{}d}t} - 2 \int_{R_c}^{R_I(t)} \rho{}(r, t)
\frac{\partial{}\rho{}(r,t)}{\partial{}t} r^2 {\rm{}d}r \right],
\label{equation:ionization_front_radius}
\end{multline}
where we assumed the ionization radius is a continuous function of time, and
used the general form of Leibniz' rule:
\begin{multline}
\frac{{\rm{}d}}{{\rm{}d}x} \left( \int_{A}^{b(x)} f(x,y) {\rm{}d}t \right) = \\
f(x,b(x)) \frac{{\rm{}d}}{{\rm{}d}x}b(x) +
\int_{A}^{b(x)} \frac{\partial{}}{\partial{}x}f(x,t) {\rm{}d}t.
\end{multline}

\subsection{Stability analysis}
\label{subsection:stability_analysis}

For what follows, we will assume a constant source luminosity $Q(t)$, 
and assume we already have a steady state solution as derived in 
\ref{subsection:ionizing_radiation}. We will now perturb this solution 
by introducing a small density perturbation outside the ionization 
radius. As shown by \eqref{equation:ionization_front_radius}, this will 
not affect the ionization radius, and the perturbation will accrete 
until it reaches the ionization front radius. Once it reaches the 
ionization front and crosses it by a small distance $\varepsilon{}$, the 
ionization front radius will change according to
\begin{multline} 
\frac{{\rm{}d}R_I(t)}{{\rm{}d}t} = -\frac{2}{\rho{}^2(R_I(t), 
t)R_I^2(t)}\\ \int_{R_I(t) - \varepsilon{}}^{R_I(t)} 
\rho{}(r,t)\frac{\partial{}\rho{}(r,t)} {\partial{}t}r^2 
{\rm{}d}r,\label{equation:perturbed_ionization_front}
\end{multline}
since this is the only part of the integral that is non trivial. 
This expression does not lend itself to a rigorous linear 
perturbation analysis because of the presence of an integral in a 
differential equation, so that we will restrict ourselves to a 
semi-analytic analysis below.

What happens next depends on the sign of 
${\partial{}\rho{}(r,t)}/{\partial{}t}$. If this sign is positive (a 
positive perturbation), the value of the integral will be positive and 
the ionization front will move inwards. For a negative perturbation, the 
ionization front will move outwards.

In either case, the perturbation does not cause a restoring force that 
balances out the perturbation and restores the initial ionization front 
radius, as the perturbation will be further accreted inside the ionized 
region, and will keep contributing to 
\eqref{equation:perturbed_ionization_front}. Depending on the size of 
the perturbation, the ionization radius will either settle into a new 
dynamic equilibrium value, or will keep moving in the same direction. In 
the limit of a very large positive perturbation, the ionization front 
will move with the accreting bump and collapse entirely. This 
qualitative behaviour of the ionization front radius constitutes a first 
important result: \emph{the change in ionization front caused by a 
perturbation cannot be undone while that perturbation is still present 
inside the ionized region}. The rate at which the ionization 
radius changes depends on the size and the shape of the perturbation.

Once the initial perturbation leaves the ionized region by crossing the 
cut off radius $R_c$, the situation is nominally reset to the original 
two temperature situation. However, since the profile behind the 
perturbation adjusted to a new ionization radius while the perturbation 
was still present, the amount of available material to ionize will now 
no longer match the initial, stable two temperature solution. For a 
positive perturbation, the ionization front will have shrunk and the 
density will now be too low; for a negative perturbation the reverse 
will have happened. We now end up in the opposite scenario of what we 
started out with: the initially positive perturbation now results in a 
profile with a negative perturbation, while the initially negative 
perturbation is now a positive perturbation. This is a second important 
result: \emph{once a perturbation leaves the ionized region, a new 
perturbation of opposite sign is created}.

The new perturbation of opposite sign has to ensure the ionizing 
luminosity $Q(t)$ is exactly balanced by recombinations (see 
\eqref{equation:ionization_balance}). Since the new perturbation will be 
created at a larger radius than the original perturbation, and since the 
Bondi accretion profile for the density scales as $\sim{}r^{-1.4}$, a 
larger density perturbation is needed to achieve this. This leads to the 
final important result: \emph{the newly created perturbation will always 
be larger than the original perturbation}.

In conclusion, any initially small perturbation will lead to an 
oscillation of the ionization front radius on a time scale of the free 
fall time (the time a perturbation created at the ionization front needs 
to reach the inner cut off radius) and with a growing amplitude, until the
amplitude is large enough to cause the ionization front to move at the same
speed as the perturbation.
Physically, there are four interesting scenarios, depending on the
sign and the size of the initial perturbation:
\begin{enumerate}
  \item{} A large positive perturbation will cause a collapse of the ionization
  front whereby the ionization front gets trapped inside the perturbation and
  collapses onto the central cut off radius.
  \item{} A small positive perturbation will initially cause a collapse of the
  ionization front, but the ionization front will move at a slower pace than the
  accretion velocity of the perturbation. When the perturbation reaches the
  central cut off radius, the ionization front will go through a phase of
  expansion, followed by one or more subsequent oscillations with growing
  ionization front speed, until the ionization front gets trapped inside a
  perturbation.
  \item{} A large negative perturbation will initially cause an expansion of the
  ionization front until the perturbation reaches the central cut off radius.
  At this point the ionization front will collapse as in the case of a large
  positive perturbation.
  \item{} A small negative perturbation will initially cause an expansion of
  the ionization front, followed by multiple oscillations with growing
  ionization front speed, until the ionizatio front gets trapped inside a
  perturbation.
\end{enumerate}

In the next section, we will test our semi-analytic stability analysis 
by numerically seeding known perturbations of a given sign and size that 
correspond to these four scenarios.

\section{Numerical simulations}

\subsection{1D code}

\subsubsection{Algorithm}

We use a textbook \citep{2009Toro} implementation of a 1D spherically symmetric
second order finite volume method\footnote{Available from
\url{https://github.com/bwvdnbro/HydroCodeSpherical1D}}.
We use a very conservative version of the
slope limiter of \citet{2015Hopkins}. Fluxes are estimated using an HLLC Riemann
solver \citep{2009Toro} with a polytropic index $\gamma{}=1.001$ (we tested our
results against an exact Riemann solver), and are limited using the per-face
slope limiter of \citet{2015Hopkins}. After every time step the pressure is
reset using the isothermal equation of state, effectively mimicking the use of
an isothermal Riemann solver. The spherical source terms needed to make the
method spherically symmetric are added using a second order Runge-Kutta method,
as suggested by \citet{2009Toro}.

Gravity is added as an extra source term in the momentum equation, using an
operator splitting method that uses a leapfrog scheme \citep{2010Springel}. We
assume that the central mass is much larger than the total mass of the fluid,
such that we only consider the mass of the central object, as an external force.

We know from \sectionref{section:analytics} that the central density is
divergent, which will cause our numerical method to break down for small radii.
We therefore set up an inner boundary radius with outflow boundary conditions,
inside which we do not follow the hydrodynamics. At a much larger radius, we set
up an inflow boundary where we impose the constant accretion rate.

To include ionization, we assume the Str\"{o}mgren approximation introduced
above, so that the ionization radius is determined by finding the root of the
following function:
\begin{multline}
f_{\rm{}ion}(R_I) = \\
\sum_{j, r_j < R_I} \rho{}_j^2 \frac{1}{3} \left[ \left(r_j
+ \frac{1}{2} \Delta{} r\right) - \left(r_j - \frac{1}{2} \Delta{} r\right)
\right]^3 - Q',
\label{equation:numerical_ionization}
\end{multline}
where $r_j$ is the central radius of cell $j$; $\rho{}_j$ is its density.
$\Delta{}r$ is the radial size of a single cell, while
\begin{equation}
Q' = \frac{m_{\rm{}H}^2 Q}{4\pi{} \alpha{}_B(T_i)} - Q_{\rm{}inner}
\end{equation}
is the ionizing luminosity that
makes it out of the inner mask (we will treat $Q'$ as a simulation parameter).

Once the ionization radius is found, we set the pressure of the region based on
the following adapted isothermal equation of state:
\begin{equation}
P(r) = \rho{}(r) \begin{cases}
C_s^2 & r >= R_I,\\
C_s^2 P_c & r < R_I,
\end{cases}
\end{equation}
where $P_c$ is the pressure contrast between the ionized and neutral region that
was introduced in \sectionref{section:analytics}.

\subsubsection{Simulation parameters}

We want to set up physically plausible initial conditions for which our 
assumptions hold, i.e. we want the density to be high enough to 
guarantee $t_{\rm{}rec} << t_{\rm{}dyn}$, while at the same time having 
a density that is low enough to actually be ionized by a physically 
plausible input luminosity $Q'$. For a central stellar mass of 
$18~{\rm{}M}_\odot{}$, we expect an ionizing luminosity 
$Q_i\sim{}10^{48}~{\rm{}s}^{-1}$ \citep{2003Keto, 
2003Sternberg}. For the steady state solution, the ionization balance 
equation is given by
\begin{equation}
Q_i = \frac{4 \pi{} \alpha{}_i \rho{}_i^2(R_I) R_I^4 
v_i^2(R_I)}{m_{\rm{}H}^2} \int_{R_c}^{R_I} \frac{1}{r^2 v_i^2(r)} 
{\rm{}d}r.
\end{equation}
We have to numerically integrate this equation to find the ionizing 
luminosity that corresponds to a given set of two temperature Bondi 
parameters.

\begin{figure}
\centering{}
\includegraphics[width=0.48\textwidth]{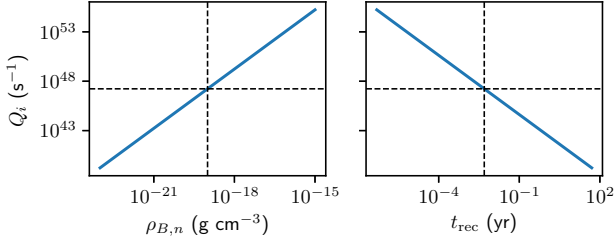}
\caption{Ionizing luminosity as a function of neutral Bondi density
(\emph{top}), and as a function of ionization front recombination time
(\emph{bottom}) for a fixed central mass, pressure contrast, neutral sound speed
and ionization front radius, assuming an inner cut off radius $R_c =
10~{\rm{}AU}$ and an ionization front radius $R_I = 30~{\rm{}AU}$. The dashed
lines correspond to the specific choice of $\rho{}_{B,n}$ we will use for
our simulations.
\label{figure:Q_vs_rho}}
\end{figure}

\figureref{figure:Q_vs_rho} shows the ionizing luminosity $Q_i$ as a 
function of the neutral Bondi density $\rho{}_{B,n}$, for a central mass 
$M = 18~{\rm{}M}_\odot{}$, neutral sound speed $C_{s,n} = 
2.031~{\rm{}km~s}^{-1}$ (corresponding to a hydrogen only gas with $T_n 
= 500~{\rm{}K}$), pressure contrast $P_c = 32$ (corresponding to an 
ionized temperature $T_i = 8000~{\rm{}K}$), and assuming an ionization 
front radius $R_I = 30~{\rm{}AU}$ (similar to Lund \emph{et al.}, 
\emph{submitted to MNRAS}) and inner cut off radius $R_c = 10~{\rm{}AU}$ 
(due to numerical precision issues, we cannot compute the argument to 
the Lambert W function for radii $< \sim{}10~{\rm{}AU}$). Also shown is 
the corresponding recombination time scale at the ionization front, 
computed as $T_d = {m_{\rm{}H}}/{(\rho{}_n(R_I) \alpha{}_i)}$, with 
$\alpha{}_i = 4\times{}10^{-13}~{\rm{}cm}^3~{\rm{}s}^{-1}$. For 
realistic ionizing luminosities $\sim{}10^{48}~{\rm{}s}^{-1}$, we find a 
neutral Bondi density $\sim{}10^{-19}~{\rm{}g~cm}^{-3}$, and a 
corresponding recombination time scale $\sim{}10^{-2}~{\rm{}yr}$. At the 
ionization radius, the free fall time is $T_{ff} = \sqrt{{2 
R_I^3}/{(GM)}} = 8.723~{\rm{}yr}$ (the free fall time taking into 
account the Bondi velocity as initial velocity is of the same order of 
magnitude), so the recombination time is more than a factor 100 smaller 
than the typical dynamical time in the entire ionized region.

We also tested this assumption using an expensive 1D time dependent 
Monte Carlo radiation transfer algorithm (Tootill, \emph{private comm.}) 
and found an excellent agreement between the time dependent results and 
the results obtained by assuming instanteneous ionization.

In what follows, we will hence use a neutral Bondi density $\rho_{B,n} =
10^{-19}~{\rm{}g~cm}^{-3}$. We will use a simulation box that spans from $R_c =
10~{\rm{}AU}$ to $R_m = 100~{\rm{}AU}$, subdivided into 2700 equal sized radial
cells, such that our desired ionization radius is on a cell boundary.

\subsubsection{Stable solution}

As a first test of the 1D code, we will try to obtain the stable two temperature
solution derived in \sectionref{section:analytics}, starting from a constant
density and velocity, and setting a constant ionization front radius $R_I =
30~{\rm{}AU}$. The analytic solution is entirely fixed by specifying the neutral
Bondi parameters, ionization front radius and pressure contrast; similarly the
numerical solution should be entirely determined by the inflow boundary
conditions at $R_m$, the mass of the external point mass, and the ionization
front position and pressure contrast if we do not restrict the solution at $R_c$
(which we can effectively do by using open boundaries). As initial conditions,
we hence set $\rho{}(r) = \rho{}_n(R_m)$ and $v(r) = v_n(R_m)$ everywhere.

\begin{figure}
\centering{}
\includegraphics[width=0.48\textwidth]{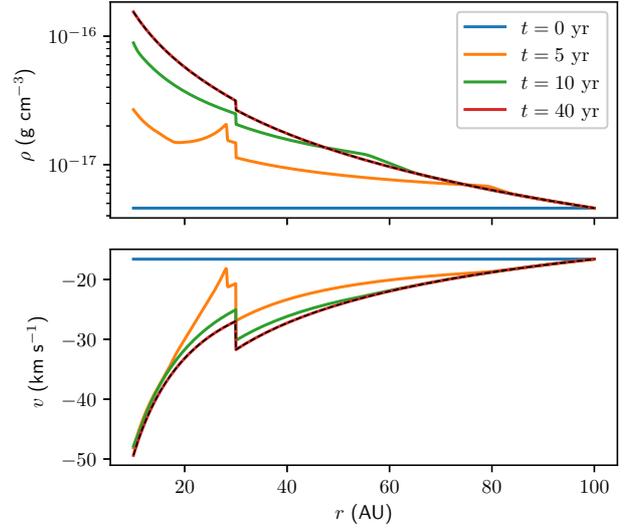}
\caption{Density and velocity profile for the steady state solution set up run
at 4 different times throughout the simulation, as indicated in the legend. The
dashed line shows the analytic two temperature solution for the same parameters.
\label{figure:stable_solution}}
\end{figure}

\begin{figure}
\centering{}
\includegraphics[width=0.48\textwidth]{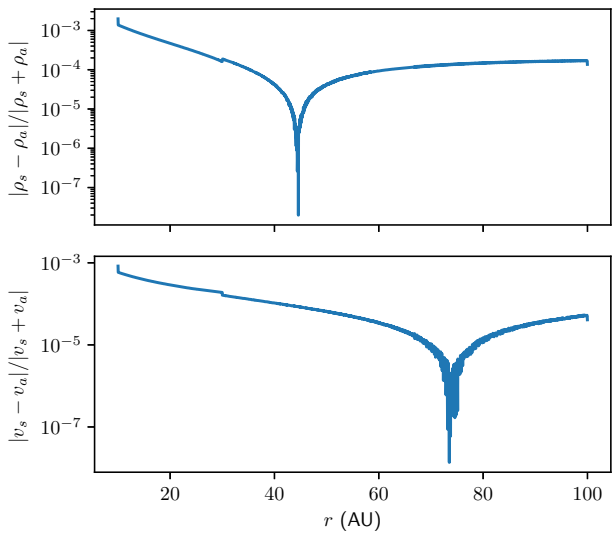}
\caption{Relative difference between the simulation and the analytic solution
for the two temperature Bondi accretion with ionization and a constant
ionization front radius, at the final simulation time $t = 40~{\rm{}yr}$.
\label{figure:stable_solution_reldiff}}
\end{figure}

\figureref{figure:stable_solution} shows the density and velocity at 4 different
times during the simulation. By $t = 20~{\rm{}yr}$, the simulation has settled
into a steady state solution. \figureref{figure:stable_solution_reldiff} shows
the relative difference between this steady state solution and the analytic
steady state solution found in \sectionref{section:analytics}, at time
$t=40~{\rm{}yr}$. Both solutions are in excellent agreement. We conclude that
our 1D code is capable of resolving the necessary physics to study this problem.

\subsubsection{Stability test}

As a second test, we test the stability of the steady state solution when the
ionization front radius is computed self-consistently. To this end, we use the
steady state output of the previous test, and compute the ionization radius
based on equation (\ref{equation:numerical_ionization}).

\begin{figure}
\centering{}
\includegraphics[width=0.48\textwidth]{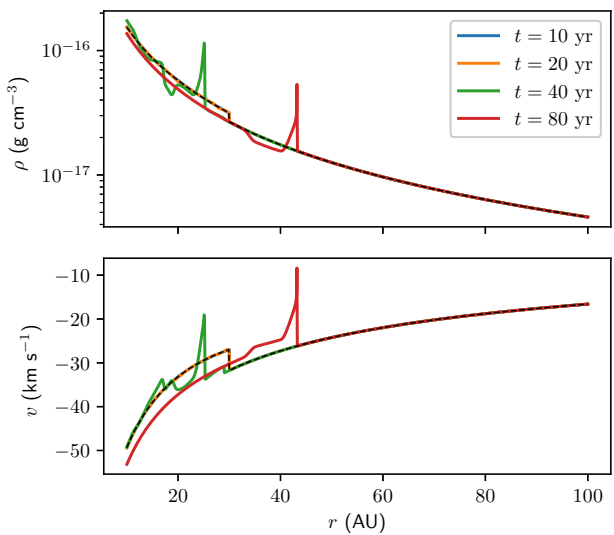}
\caption{Density and velocity profile for the stability test run at 4 different
times throughout the simulation, as indicated in the legend. The dashed line
shows the analytic two temperature solution for the same parameters.
\label{figure:instability}}
\end{figure}

\begin{figure}
\centering{}
\includegraphics[width=0.48\textwidth]{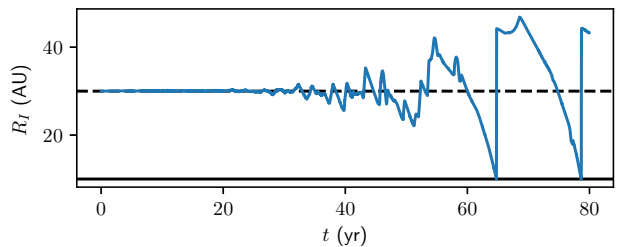}
\caption{Ionization front radius as a function of time for the stability test
run with self-consistent ionization. The dashed horizontal line shows the
initial ionization front radius, while the full horizontal line shows the
inner outflow boundary radius.
\label{figure:instability_radius}}
\end{figure}

\figureref{figure:instability} shows the resulting density and velocity profiles
at four different times during the simulation. Initially, the steady state
solution is stable. However, after $\sim{}10~{\rm{}yr}$, the density around the
ionization front starts to break up into small peaks that travel inwards with
the accretion flow. These small peaks cause tiny fluctuations in the ionization
front radius, which gradually grow. After $\sim{}20~{\rm{}yr}$, the ionization
front radius is no longer stable, and starts to oscillate, as can been seen from
\figureref{figure:instability_radius}.

It is important to note that we did not seed the small oscillations that lead to
the collapse of the steady state solution; these oscillations are seeded
numerically by accumulated round off error. We conclude that the steady state
solution is only marginally stable, as even very small scale noise can cause it
to become unstable.

\subsubsection{Convergence}

The result for the stability test discussed above depends on numerically 
seeded noise, which depends on resolution. This makes it impossible to 
perform a standard convergence test to check whether the time evolution 
of the ionization front corresponds to a physical effect or is simply 
caused by numerical issues. To show that our results do converge, we 
have to modify our set up so that we can obtain results that are 
independent of the spatial resolution for a sufficiently high spatial 
resolution. We identify two issues that need to be addressed: (a) we 
need to seed the initial collapse of the ionization front with a known 
perturbation that is the same for all resolutions, (b) we need to make 
sure that newly created instabilities that cause the recurring collapse 
of the ionization front are consistent between resolutions. Since these 
depend on the shape of the ionization front, we need to make sure we 
resolve the ionization front consistently between resolutions.

To obtain a spatially resolved ionization front, we convert 
the strong jump into a smooth linear transition. As before, we compute 
the ionization front radius $R_I$ from equation 
(\ref{equation:numerical_ionization}). However, now the neutral fraction 
$x_{\rm{}H}(r)$ of the gas does not just jump from $0$ to $1$ across the 
ionization front, but is given by
\begin{equation}
x_{\rm{}H}(r) = \begin{cases}
0 & r < R_I - \frac{3}{4S}, \\
x_{{\rm{}H},t}(r) & R_I - \frac{3}{4S} \leq{} r \leq{} R_I + \frac{3}{4S}, \\
1 & r > R_I + \frac{3}{4S}, \\
\end{cases}
\end{equation}
where $S$ is the maximal slope of the transition, and the transition width is
$W = {3}/{(2S)}$. The transition itself is a third order polynomial:
\begin{equation}
x_{{\rm{}H},t}(r) = -\frac{16}{27}S^3 (r - R_I)^3 + S (r - R_I) + \frac{1}{2}.
\end{equation}
This particular transition shape was chosen because it allows us to control the
slope of the transition, while ensuring continuous first derivatives in the
transitions points $r = R_I - W/2$ and $r = R_I + W/2$.

The adapted equation of state now becomes
\begin{equation}
P(r) = \rho{}(r) C_s^2 \left[ (1 - x_{\rm{H}}(r))P_c + x_{\rm{}H}(r) \right].
\end{equation}

\begin{figure}
\centering{}
\includegraphics[width=0.48\textwidth]{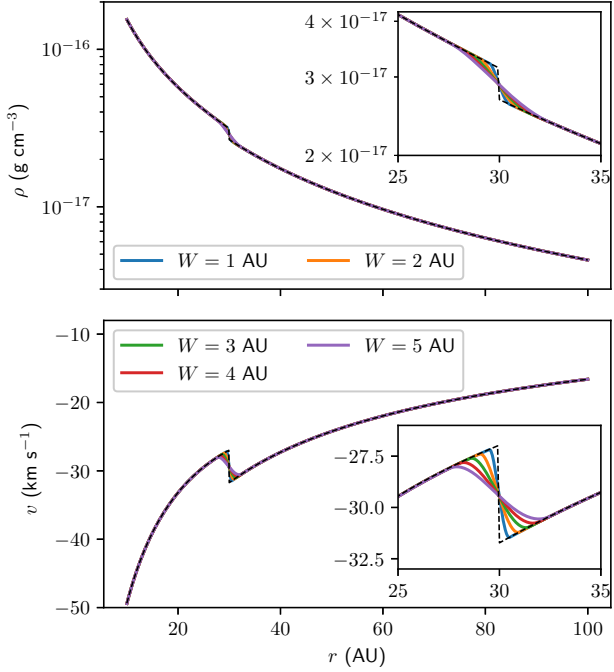}
\caption{Steady state density and velocity as a function of radius for
simulations with a smooth transition from ionized to neutral. The full lines
show 2700 cell simulation results, with the width of the corresponding smooth
transition indicated in the legend. The dashed line is the analytic two
temperature solution. The insets show a zoom of the region around the
ionization front radius, where the smooth simulations deviate from the analytic
solution.
\label{figure:convergence_stable}}
\end{figure}

\begin{figure}
\centering{}
\includegraphics[width=0.48\textwidth]{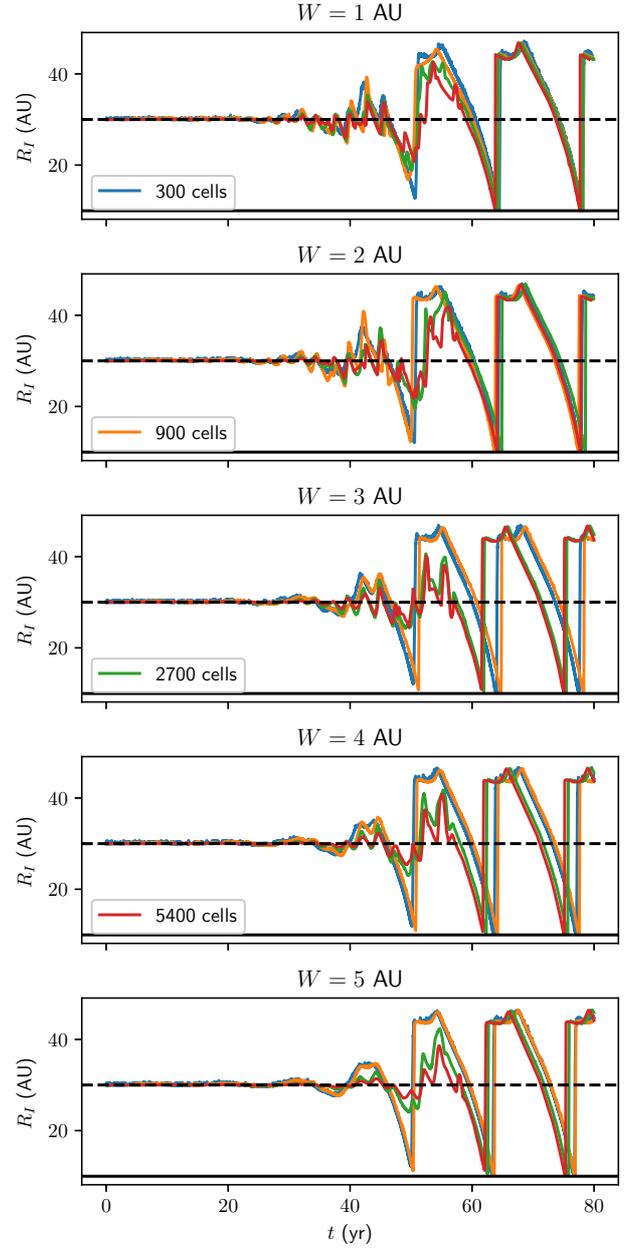}
\caption{Ionization front radius as a function of time for simulations with a
smooth transition from ionized to neutral region, self-consistent ionization,
and a numerically seeded instability. The full lines show simulation results for
different spatial resolutions, as indicated in the legend. The dashed
horizontal line shows the initial ionization front radius, the full horizontal
line shows the radius of the inner outflow boundary.
\label{figure:convergence_instability}}
\end{figure}

\figureref{figure:convergence_stable} shows the stable two temperature 
solution for a range of simulations with different values of the 
transition width $W$ and for different spatial resolutions. As expected, 
the transition region smooths out the jump from the neutral to the 
ionized region, but outside the transition region the density and 
velocity still follow the analytic solution.

\figureref{figure:convergence_instability} shows the ionization front 
radius as a function of time for the same simulations when we 
self-consistently update the ionization state. All simulations still 
become unstable after some time, with the instability still being seeded 
by numerical noise. Since the noise depends on the resolution, the 
instability is seeded differently for different resolutions, and we do 
not obtain a converged result.

To seed the instability, we use the output of the set up runs shown in 
\figureref{figure:convergence_stable} and add a small gaussian-like 
density perturbation filter $\phi{}(r)$ to the density, such that 
$\rho{}'(r) = \phi{}(r) \rho{}(r)$:
\begin{equation}
\phi{}(r) = 
\begin{cases} 1 + \rho{}_p \left( 1 - 6 u^2 + 6 u^3 \right) & u < 
\frac{1}{2}, \\ 1 + 2 \rho{}_p \left( 1 - u \right)^3 & \frac{1}{2} 
\leq{} u < 1, \\ 1 & 1 \leq{} u, \end{cases}
\end{equation}
with $u = {|r - r_0|}/{h}$, where $r_0$ is the center of the gaussian-like 
bump and $h$ its width. $\rho{}_p$ is the amplitude of the perturbation, 
and the shape of the perturbation is based on the cubic spline kernel 
commonly used in SPH codes \citep[see e.g.][]{2005Springel_Gadget}. We will use 
$r_0 = 65~{\rm{}AU}$ and $h = 5~{\rm{}AU}$. We will choose 4 different 
amplitudes for the perturbation to test the various scenarios derived 
above: $\rho{}_p = 0.1$, $\rho{}_p = 2$, $\rho{}_p = -0.1$ and
$\rho{}_p = -0.01$. These 
correspond respectively to a small positive seed that should cause an 
ionization front movement slower than the accretion velocity, a large 
positive seed that should cause the immediate collapse of the ionization 
front, a negative seed that should initially cause an expansion of 
the ionization front followed by a collapse, and a negative seed that should
cause a number of oscillations.

\begin{figure}
\centering{}
\includegraphics[width=0.48\textwidth]{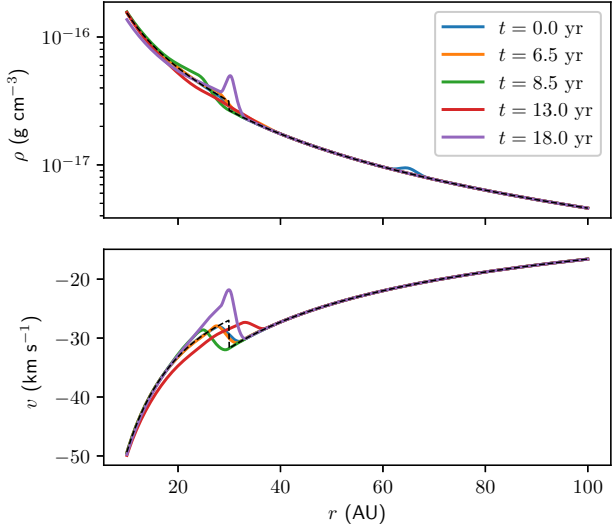}
\caption{Density and velocity profile for the stability test run with
$\rho{}_p=0.1$, at 5 significant times throughout the simulation, as indicated
in the legend. The dashed line shows the analytic two temperature solution for
the same parameters.
\label{figure:convergence_seed}}
\end{figure}

\begin{figure}
\centering{}
\includegraphics[width=0.48\textwidth]{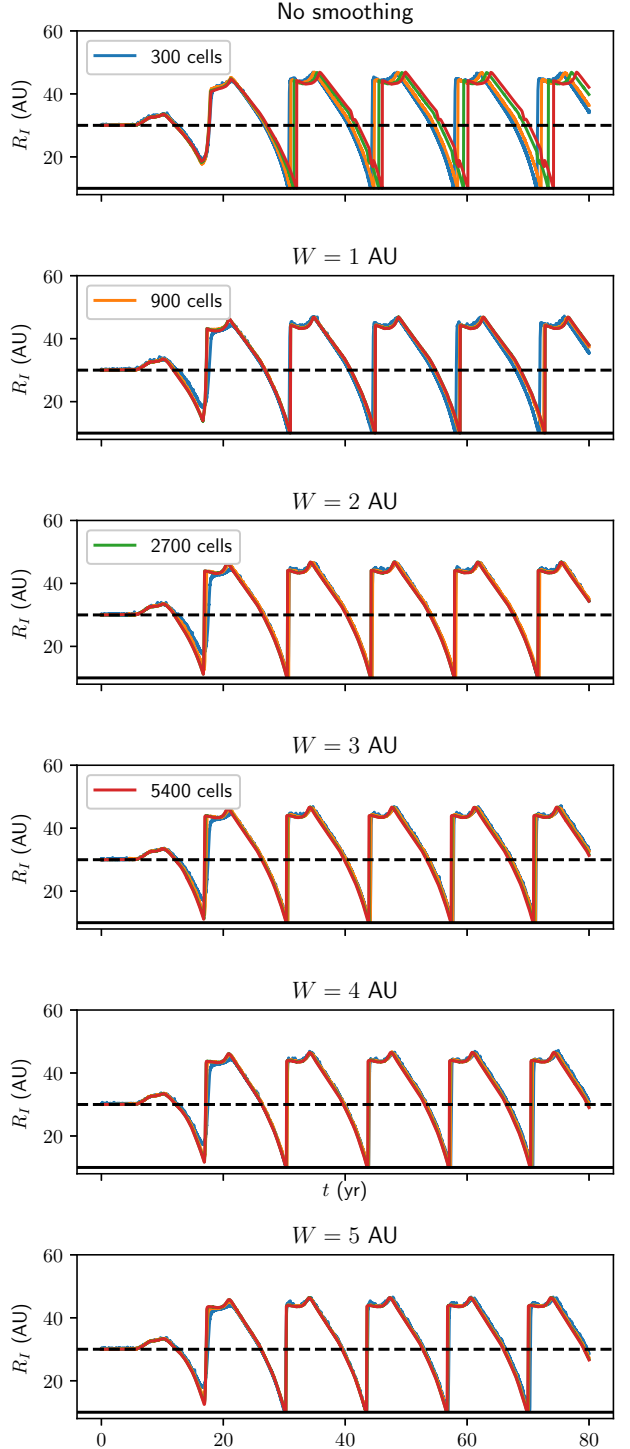}
\caption{Ionization front radius as a function of time for simulations with and without a
smooth transition from ionized to neutral region, self-consistent ionization,
and a seeded instability with $\rho{}_p=-0.1$. The full lines show simulation
results for different spatial resolutions, as indicated in the legend. The
dashed line shows the initial ionization front radius, the full line shows the
radius of the inner outflow boundary.
\label{figure:convergence_seed_radius}}
\end{figure}

\begin{figure}
\centering{}
\includegraphics[width=0.48\textwidth]{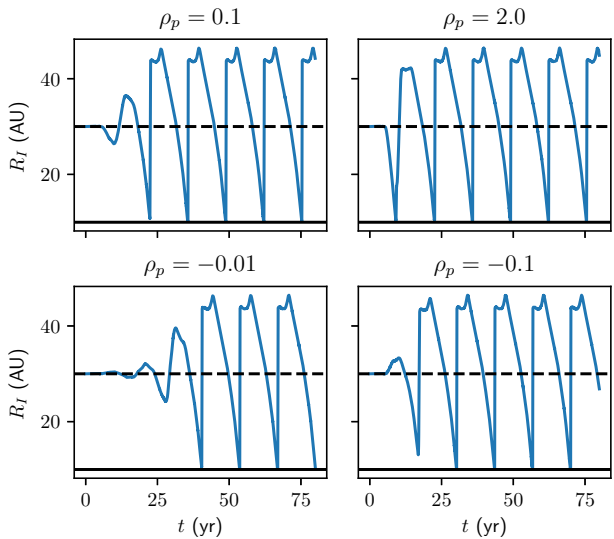}
\caption{Ionization radius as a function of time for simulations with 4
different values for the perturbation amplitude $\rho{}_p$, as indicated above
each panel. All simulations are based on the same initial condition with a
smooth ionization transition of $5~{\rm{}AU}$ and use a resolution of 2700
cells. The dashed black lines show the initial ionisation radius, while the full
black lines correspond to the radius of the inner outflow boundary.
\label{figure:convergence_seed_radius_all}}
\end{figure}

\figureref{figure:convergence_seed} shows the time evolution of a 
simulation with a small positive seed perturbation. The perturbation 
initially moves inwards with the accretion flow until it reaches the 
ionization front. Once it arrives there, it causes an increase in the 
density inside the ionization region, which causes the ionization front 
radius to shrink according to equation 
(\ref{equation:ionization_front_radius}). The density enhancement 
continues to move inwards with the accretion flow, and for a while the 
ionization front moves inwards as well. The region outside the 
ionization front radius in the meantime adjusts to the lower pressure 
and settles into a neutral Bondi accretion, with an overall lower 
density.

When the initial perturbation reaches the inner outflow boundary, the 
mass inside the perturbation is effectively lost from the system, 
causing the density inside the ionized region to go down again, and 
subsequently the ionization front radius to expand again. However, since 
the density inside the original ionization front radius partially 
settled into a neutral Bondi profile, the overall density inside the 
ionized region will be lower than in the steady state two temperature 
solution, and the new ionization front radius will expand beyond the 
original ionization front radius. This is not a steady state solution 
anymore, so the ionization front radius will consequently shrink again 
as the inner density adapts to the ionization again. During that 
process, a large peak in density builds up around the ionization front 
radius, which in combination with the ionization front evolution 
(equation (\ref{equation:ionization_front_radius}) leads to a runaway 
collapse of the ionization front towards the mask. Once this peak 
reaches the inner outflow boundary, the situation is reset again, and 
the whole process repeats itself.

\figureref{figure:convergence_seed_radius} shows the evolution of the 
ionization front radius as a function of time for the simulations with a 
negative seed perturbation with $\rho{}_p = -0.1$. We clearly see how 
the repeated runaway collapse of the ionization front leads to a 
periodic oscillation of the ionization front radius with a period of 
$\sim{}10~{\rm{}yr}$, roughly consistent with the free fall time of the 
system. It is clear that the combination of a smooth transition that 
damps small scale noise and the seeding of the instability with a well 
defined perturbation allow us to achieve numerical convergence for the 
high resolution simulations. This in principle should allow us to use 
this test as a benchmark for this type of instability. Note that the 
convergence is better for initial conditions with a larger transition 
region from fully ionized to fully neutral. For what follows, we will 
hence limit ourselves to simulations with an ionization transition of 
$5~{\rm{}AU}$ and a resolution of 2700 cells.

Note that the expanding ionization front at the end of each oscillation 
moves relatively fast, so that we cannot guarantee that the dynamical 
time for this expansion is significantly larger than the recombination 
time; a main assumption for our ionization treatment. However, the 
overall density inside the central region is low enough to guarantee the 
build up of a new ionization front at the expected radius in a time that 
is short enough not to significantly alter our dynamics, and we can 
hence still assume our results are qualitatively correct, albeit with a 
slightly longer oscillation period. We confirmed this using a time dependent
Monte Carlo radiative transfer simulation. Similarly, moving the inner outflow 
boundary to smaller radii will lead to a small increase in oscillation 
period, but will not change the qualitative behaviour, as we still 
expect accreted material at small enough radii to be lost from our 
system of interest.

\figureref{figure:convergence_seed_radius_all} shows the time evolution 
of the ionization radius for the 4 different scenarios discussed in 
\ref{subsection:stability_analysis}. In the case of a positive density 
perturbation, the ionization front initially collapses. If the bump is 
small, the ionization front does not move at the same speed as the bump, 
and the collapse halts when the bump moves accross the inner outflow 
boundary. If the bump is large enough, the ionization front moves along 
with the bump and both collapse onto the inner outflow boundary.

For a negative density bump, the ionization front initially expands 
until the density bump is accreted accross the inner boundary. At this 
point, the expanded ionization region will start to contract again. For 
a large negative bump, this immediately leads to a collapse of the 
ionization front. For smaller bumps, the collapse is only partial and 
leads to an oscillation of the ionization front with an amplitude that 
increases over time.

In all cases, the initial perturbation eventually evolves into a periodic
oscillation of the ionization front, i.e. all scenarios naturally evolve into
the case where the ionization front movement becomes coupled to the movement of
the density perturbation.

These simulations hence confirm the semi-analytic model we derived before.

\subsubsection{TORUS comparison}

To investigate the impact of our approximations on the result of the 1D 
models, we reran our 1D test with the radiation hydrodynamics code 
\textsc{torus} \citep{2000Harries}, using the ``detailed'' model as 
described in \citet{2015Haworth}. This includes polychromatic radiation 
transport, diffuse field radiation (i.e. not the on-the-spot 
approximation) as well as helium, metals and the associated line cooling 
in the thermal balance. In these calculations a hydrodynamics-only 
calculation is run until a steady state is reached. The medium is the 
fully ionized, which lowers the optical depth and hence speeds up the 
convergence of the first photoionization equilibrium calculation. 
Subsequently, photoionization and hydrodynamics steps are performed 
iteratively to follow the RHD evolution. The ionizing luminosity of the 
source was tuned to approximately reproduce the same ionization front 
radius as in the models above.

\begin{figure}
\centering{}
\includegraphics[width=0.48\textwidth]{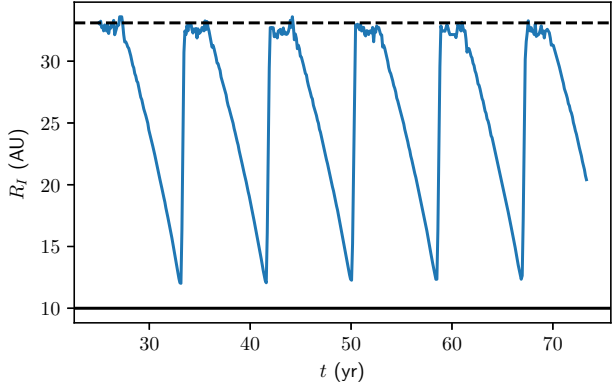}
\caption{Ionization front as a function of time for a self-consistent RHD
simulation of a single ionizing source within an initially stable neutral
Bondi profile, simulated with \textsc{torus}. The black lines represent the
initial ionisation front radius (dashed) and the mask radius (full).
\label{figure:torus}}
\end{figure}

The resulting time evolution of the ionization front is shown in 
\figureref{figure:torus}. As in the models above, the ionization front 
periodically shrinks and then reappears. This shows that even with a 
more detailed treatment of the photoionization, perturbations of the 
two-temperature solution are enhanced by the spherical symmetry of the 
system and never dampen out. It is interesting to note that in this 
case, the ionisation front does not collapse all the way up to the mask. 
The reason for this is that the ionization front in this case is located 
upstream from the density peak that causes the collapse, so that the 
density peak is absorbed by the mask before the ionization front reaches 
the mask radius.

\subsection{3D simulation}

\begin{figure}
\centering{}
\includegraphics[width=0.48\textwidth]{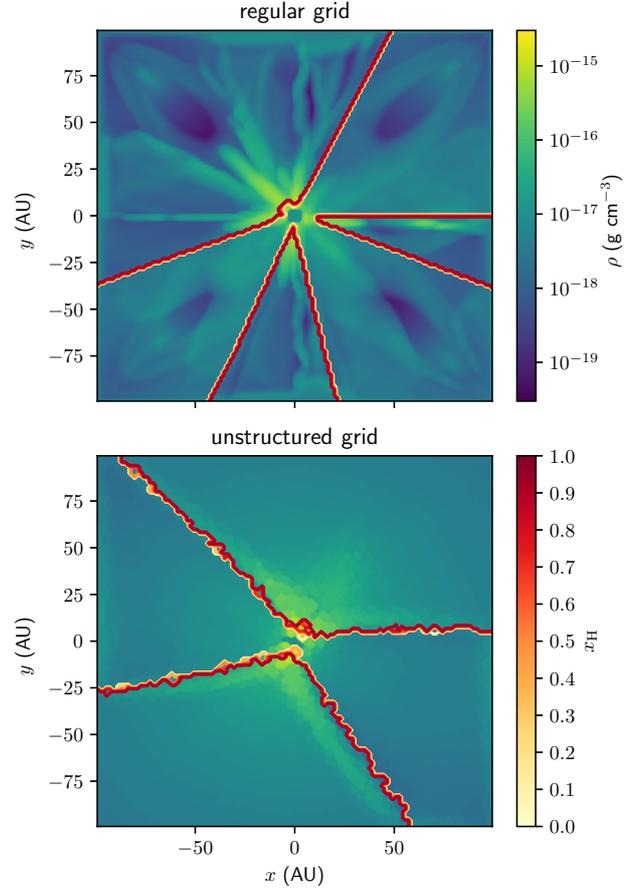}
\caption{Density slice at $z=0$ for a 3D Bondi accretion
simulation with self-consistent photoionization and using a regular Cartesian
grid of $128\times{}128\times{}128$ cells (\emph{top}) and an unstructured
Voronoi grid with 100,000 cells (\emph{bottom}). Both simulations were started
from the same initial density profile and evolved self-consistently until
$t = 20~{\rm{}yr}$.
\label{figure:3D_bondi}}
\end{figure}

The presence of a 1D instability has important implications for 3D RHD
simulations of spherically symmetric Bondi accretion. If spherical symmetry is
not explicitly imposed on the radiation field or on the coupling between
radiation field and hydrodynamics, then the instability will be seeded
differently in different directions. Once initial asymetries have been seeded,
the problem is no longer spherically symmetric and true 3D effects will govern
the hydrodynamical evolution.

We explore these effects by running the same Bondi set up described 
above in a full 3D RHD simulation using the Monte Carlo RHD code 
\textsc{CMacIonize} \citep{2018Vandenbroucke_CMacIonize}. Since we are not 
interested in the detailed results, we limit ourselves to low resolution 
simulations and do not explicitly check for convergence. More realistic 
simulations will be subject of future studies. We run two simulations 
with different grids used to discretize the fluid. Both simulations 
follow a fluid in a box of 
$100~{\rm{}AU}\times{}100~{\rm{}AU}\times{}100~{\rm{}AU}$ with inflow 
boundary conditions and a point mass (and ionization source) at the 
centre. Just as for the 1D simulations, a central sphere with radius 
$10~{\rm{}AU}$ is masked out. The mass, initial ionization radius, 
neutral Bondi density, neutral gas temperature and pressure contrast are 
set as before.

A first set up uses a regular Cartesian grid of 
$128\times{}128\times{}128$ cells. To compute the ionization front 
position, we use $10^7$ Monte Carlo photon packets and 10 iterations. 
Initial instabilities will be seeded by two mechanisms: Poisson noise 
inherent to the Monte Carlo method, and discretization error due to the 
low resolution grid used. The latter will have a strong dependence on 
specific grid directions (e.g. the coordinate axis directions and 
diagonals), and from the top panel of \figureref{figure:3D_bondi} it 
is clear that they constitute the dominant seeding mechanism, resulting 
in a very symmetrical, yet no longer spherically symmetric ionization 
region.

The higher pressure in the directions with a larger ionization radius 
will cause a tangential force that evacuates these regions and pushes 
material into directions that are still neutral, reenforcing the 
differences between both regions. While some directions with larger 
ionization radii still periodically collapse due to the radial 
instability, the directions with a smaller ionization radius actually 
remain stable. At later times, the ionization region has a chaotic, 
asymmetric shape, while the density field shows clear signs of grid 
symmetries.

To eliminate the effect of Cartesian grid symmetries, we run the same 
simulation using an unstructured Voronoi grid consisting of 100,000 
cells, using $10^6$ photon packets for 10 iterations. This is the same 
grid structure used to run \textsc{CMacIonize} as a moving-mesh code, 
but without actually moving the mesh in between time steps (since this 
would be very complicated for a 3D problem with spherical accretion). In 
this case, the results look less artificial (see bottom panel of 
\figureref{figure:3D_bondi}), but the qualitative interpretation is 
the same: the combination of the 1D instability and 3D effects will 
cause the ionization front to break up into a non-spherically symmetric 
structure, and the ionization front will stabilize in some directions.

We conclude that unless ionization is treated in a perfectly spherically 
symmetric manner, 3D simulations will never result in a perfectly 
symmetric ionization region, irrespective of any additional 3D effects 
that could be included in these simulations. This is valuable background 
knowledge for the interpretation of future 3D studies of Bondi accretion 
that will include rotation, and a general limitation for RHD codes in 
studying problems that involve both accretion and photoionization.

\section{Conclusion}

In this work, we studied the stability of spherically symmetric ionized 
Bondi accretion flows around young stellar objects. We derived an 
anlytic expression for a two temperature steady state solution in which 
an R type ionization front with a size of the order of $10~{\rm{}AU}$ is 
trapped in the highly supersonic accreting flow, consistent with the 
predictions of \citet{2002Keto, 2003Keto}. We were able to accurately 
reproduce this analytic profile in a constrained 1D RHD simulation.

However, we also showed that this steady state solution is only 
marginally stable if ionization is treated self-consistently, leading to 
a collapse of the ionization front under small perturbations in density 
or ionizing luminosity. We were able to confirm our semi-analytic model 
of this instability using 1D RHD simulations, and were able to obtain 
spatially converged results by introducing a noise suppressing 
transition layer and a seeded instability.

Both the constrained simulations that lead to the two temperature steady 
state solution and the converged simulations of the instability could 
serve as benchmark problems for 1D RHD codes.

Finally, we showed that the existence of a 1D instability has profound 
implications for 3D simulations of ionized accretion flows when the 
radiation field is not treated in a perfectly spherically symmetric 
manner. The instability makes it impossible to retain spherical 
symmetry, and will amplify grid symmetries.

\section*{Acknowledgements}
We want to thank the anonymous referee for constructive and insightful 
comments regarding the convergence of our simulations. BV and KW 
acknowledge support from STFC grant ST/M001296/1. NS would like to thank 
CAPES for graduate research funding. KL acknowledges support from the 
Carnegie Trust. DFG thanks the Brazilian agencies FAPESP (no. 
2013/10559-3) and CNPq (no. 311128/2017-3) for financial support. TJH is 
funded by an Imperial College London Junior Research Fellowship.

\bibliographystyle{mnras}
\bibliography{main}

\appendix{}

\section{Analysis of strong and weak R type solutions}
\label{appendix:Gamma}

Here we perform a detailed analysis of equation (\ref{eq:Gamma}) for the case of
an R type front, with $v_n(R_I) < -v_R$. We begin by rewriting the equation in
terms of the dimensionless variables $x = -{v_n(R_I)}/{C_{s,i}}$ and $A =
{C_{s,n}}/{C_{s,i}}$:
\begin{equation}
\Gamma{}(x, A) = \frac{1}{2} \left( x^2 + A^2 \pm{} \sqrt{(x^2+A^2)^2 - 4x^2}
\right).
\end{equation}
Note that $A = \sqrt{{1}/{P_c}} < 1$. To get a real solution, we require
\begin{equation}
x > 1 + \sqrt{1 - A^2},
\end{equation}
and hence the neutral region is supersonic. For the critical value $x_c = 1 +
\sqrt{1 - A^2}$,
\begin{equation}
\Gamma{}_c(A) = 1 + \sqrt{1 - A^2} = x_c > 1,
\end{equation}
which implies $v_i(R_I) = -C_{s,i} > v_n(R_I)$. We now explore how the value of
$\Gamma{}$ changes for $x>x_c$ in the case of a strong and a weak R type front.

The first derivatives of $\Gamma{}(x, A)$ are
\begin{align}
\frac{\partial{}\Gamma{}(x, A)}{\partial{}x} &= x \pm{} \frac{(A^2 + x^2
-2)x}{\sqrt{(x^2 + A^2)^2 - 4x^2}}, \\
\frac{\partial{}\Gamma{}(x, A)}{\partial{}A} &= A \pm{} \frac{(A^2 +
x^2)A}{\sqrt{(x^2 + A^2)^2 - 4x^2}}.
\end{align}
For $x > x_c$ and $0 < A < 1$, these derivatives have no real roots, which means
$\Gamma{}(x, A)$ is monotonous in these regions. Furthermore, we have that
${\partial{}\Gamma{}(x, A)}/{\partial{}x} \rightarrow{} \pm{} \infty{}$ for
$x \rightarrow{} x_c$, which means that $\Gamma{} > \Gamma{}_c$ for a strong R
front, and $\Gamma{} < \Gamma{}_c$ for a weak front.

For the weak front, we find (note the signs):
\begin{equation}
\Gamma{} = \frac{v_n(R_I)}{v_i(R_I)} < \Gamma{}_c = x_c = \frac{v_R}{C_{s,i}} <
\frac{-v_n(R_I)}{C_{s,i}},
\end{equation}
or
\begin{equation}
v_i(R_I) < -C_{s,i},
\end{equation}
which means that the ionized region is always supersonic. For a strong R type
front the condition $\Gamma{} > \Gamma{}_c$ does not impose similar
restrictions, so that a strong front can be subsonic or supersonic.

Note that there is some confusion in literature about strong and weak R type
fronts and whether or not they are supersonic or subsonic. From our analysis, it
follows that a \emph{weak} R type front, defined as the front with the smaller
jump in density and velocity across the ionization front, is \emph{always}
supersonic, and is the only front expected to exist in nature. The \emph{strong}
front, defined as the front with a larger jump in density and velocity across
the ionization front, can be either supersonic or subsonic. For the critical
case $v_n(R_I) = -v_R$, the strong and weak front are the same, and the ionized
velocity is the ionized sound speed.

\bsp    
\label{lastpage}

\end{document}